\documentclass[11pt,a4paper]{article}
\usepackage[utf8]{inputenc}
\usepackage[english]{babel}
\usepackage{amsmath}
\usepackage{amsfonts}
\usepackage{amssymb}
\usepackage{makeidx}
\usepackage{graphicx}
\usepackage{jcappub}
\usepackage{longtable}
\usepackage{multirow}
\usepackage{bigstrut}
\usepackage{hyperref}

\usepackage{multicol} 

\title{CMB low multipole alignments in the $\mathbf{\Lambda}$CDM and Dipolar models}
\author[a]{L.Polastri}
\author[b,c]{, A.Gruppuso}
\author[a,b,d]{and P. Natoli}
\affiliation[a]{Dipartimento di Fisica e Scienze della Terra, Universit\`a degli Studi di Ferrara,\\ via Giuseppe Saragat 1, I-44122 Ferrara, Italy}
\affiliation[b]{INAF-IASF Bologna,\\ Via Piero Gobetti 101, I-40129, Bologna, Italy}
\affiliation[c]{INFN, Sezione di Bologna,\\ Via Irnerio 46, I-40126 Bologna, Italy}
\affiliation[d]{INFN, Sezione di Ferrara,\\ via Giuseppe Saragat 1, I-44122 Ferrara, Italy}
\emailAdd{linda.polastri@student.unife.it}
\emailAdd{gruppuso@iasfbo.inaf.it}
\emailAdd{natoli@fe.infn.it}
\abstract{
The dipolar model \cite{Gordon:2005ai} has attracted much interest because it may phenomenologically explain the CMB hemispherical power asymmetry found in the WMAP and Planck data. Since such a model explicitly breaks isotropy at large angular scales it is natural to wonder whether it can also explain other CMB directional anomalies.
Focusing on the low $\ell$ alignments and assuming $\Lambda$CDM, we confirm that the quadrupole/octupole and the dipole/quadrupole/octupole alignments are anomalous with a significance up to $99.9\%$ C.L., for both WMAP and Planck data. Moreover, we show for the first time that such features are anomalous also in the dipolar model, roughly at the same level as in $\Lambda$CDM. We conclude that the dipolar model does not provide a better fit to the data than the $\Lambda$CDM.
}
\arxivnumber{}

\begin{document}
\maketitle

\section{Introduction}
Cosmic microwave background, henceforth CMB, anisotropy observations (as well as other astrophysical and cosmological observations) can be described with just six parameters in the $\Lambda$CDM model. 
To date, no extension of this model has improved in a significant way the fit to the available data \cite{Hinshaw:2012aka, Ade:2013zuv}. 
It is impressive that all the huge amount of data arising from cosmological observations seem to suggest that such simple model is sufficient to describe the large scale universe we live in.
However, observed features exist that are not very well explained by the $\Lambda$CDM model. 
This is the case of the largest CMB angular scales where so-called anomalies occur.
These can be grossly divided in two classes: isotropic and anisotropic anomalies. 
Examples of the former are the lack of power at large angular scale \cite{Monteserin:2007fv,Cruz:2010ud,Gruppuso:2013xba,Ade:2013nlj}, 
the lack of correlation in the two-point correlation function \cite{Copi:2006tu,Bernui:2006ft,Copi:2008hw,Gruppuso:2013dba,Copi:2013cya} 
and the so-called point-parity anomaly \cite{Land:2005jq,Kim:2010gf,Kim:2010gd,Gruppuso:2010nd,Aluri:2011wv,Ade:2013nlj}.
In the latter we list the hemispherical power asymmetry \cite{Eriksen:2003db,Hansen:2004vq,Eriksen:2007pc,Hansen:2008ym,Hoftuft:2009rq,Paci:2010wp, Paci:2013gs,Ade:2013nlj}, 
the mirror-parity anomaly \cite{BenDavid:2011fc,Finelli:2011zs,Ade:2013nlj,Ben-David:2014mea, Rassat:2014yna}, 
the cold spot \cite{Vielva:2003et,Cruz:2004ce, Cruz:2006sv, Vielva:2010ng} and the low $\ell$ alignments 
\cite{Copi:2003kt,Land:2005dq,Abramo:2006gw,Land:2007bn,Gruppuso:2007cn,Gruppuso:2009ee,Gruppuso:2010up,Copi:2010na,Copi:2013jna}.
The significance of these anomalies is in general of the order of 2-$3 \, \sigma$, rarely more.

A key point is whether these anomalies can be ascribed to residual systematic contamination (of astrophysical or instrumental origin), or may hint to new physics. 
Since we now know that the CMB anomalies are consistently observed in both WMAP and Planck data, little room is left for the possibility that they are artificially created by residual systematic effects.
The high quality level of foreground component separation performed by Planck \cite{Ade:2013hta} appears to rule out the case for residual foreground contamination unless there are unaccounted ingredient to the foreground model, see e.g. \cite{Liu:2014mpa} for a possible candidate.
The simplest explanation is that of statistical flukes; such line of reasoning is supported when properly accounting for multiplicity of tests also known
as the ``look-elsewhere effect'' \cite{Gross:2010qma}.
However, the number of these features, the fact that not all of them are related one another in an obvious manner and their almost exclusive occurrence at large angular scales 
motivate the quest for a (possibly unifying) explanation even if the individual statistical significance is not very high\footnote{Note that such significance 
is largely dominated by cosmic variance in the underlying $\Lambda$CDM model assumed.}.

In the current paper we focus on the low $\ell$ alignments, namely the unlikely alignments between the quadrupole and the octupole, as well as the dipole with both of the former. 
In the light of several foreground cleaned CMB maps released by both WMAP and Planck, 
we aim at assessing the statistical significance of these features.
In so doing, we test not only the $\Lambda$CDM model, but also the so called dipolar model. 
The latter is a phenomenological model which has been invoked to explain the already mentioned power hemispherical asymmetry \cite{Bernui:2008cr, Hansen:2008ym, Hoftuft:2009rq, Eriksen:2007pc, Bernui:2006ft, Ade:2013nlj}. 
The dipolar model \cite{Gordon:2005ai} consists of a particular mechanism for breaking the isotropy on the large-angle CMB fluctuations. 
The model is described by:
\begin{equation}
\left( \frac{\Delta T}{T} \right) _{mod} (\widehat{n}) = (1 + A \widehat{n} \cdot \widehat{p})  \left( \frac{\Delta T}{T} \right) _{iso}(\widehat{n}) ,
\label{dipolarmodel}
\end{equation}
where $\widehat{n}$ is the observed direction, $\left( \Delta T / T  \right) _{mod}$ is the observed (and modulated) CMB temperature fluctuations, 
$\left( \Delta T / T  \right) _{iso}$ is the usual isotropic CMB pattern, 
$A$ is the amplitude of the dipole modulation and $\widehat{p}$ is a given direction.
In \cite{Hoftuft:2009rq} it is found that $A = 0.07 \pm 0.022$, statistically significant at $ \sim 3 \, \sigma$ and the direction $\widehat{p}$ is given by 
$(l,b)=(224 ^{\circ},-22 ^{\circ}) \pm 22 ^{\circ} $ in Galactic coordinates, significant at $ \sim 3.3 \, \sigma$, see also \cite{Eriksen:2007pc} for previous results.

The paper is organized as follows. 
Section \ref{lowellalignment} is the bulk of this paper. In particular in Section \ref{data} we discuss the state of the art of the CMB anomalous alignments
and describe the used data set. 
In Section \ref{mv} we introduce the methodology employed, based on the multipole vectors formalism. 
We set forth the estimators adopted in Section \ref{estim} and present our 
data analysis pipeline, employed both for real data and realistic simulations in Section \ref{pipeline}. 
We present our results in Section \ref{4}
while Section \ref{conclusion} is reserved for conclusions.

\section{CMB low $\ell$ alignments}
\label{lowellalignment}

\subsection{State of the art and employed data set}
\label{data}

The occurrence of the anomalous alignments in the large angle CMB pattern has been noted since the very first appearance of the WMAP data \cite{de OliveiraCosta:2003pu}.
Using a different methodology, it was confirmed \cite{Copi:2003kt} that the quadrupole and the octupole are unlikely aligned in the WMAP ILC 1 year data (see also \cite{Weeks:2004cz} for a similar and independent analysis). 
It was later shown \cite{Copi:2006tu} that the quadrupole/octupole unlikely alignment is still present in the WMAP ILC 3 year map at $99.6\%$ C.L.. 
Moreover in the same paper a correlation between quadrupole, octupole and dipole was found with a significance of $99.7\%$.
The quadrupole/octupole alignment has also been studied in the Planck data \cite{Ade:2013nlj}, where similar conclusions were drawn although with slightly lower significance. 
The WMAP ILC 7 and 9 year maps are analyzed in \cite{Copi:2013jna} where it is reported that the quadruple/octupole alignment occurs with probability $0.327\%$ and $0.511\%$, respectively. 
In the same paper, it has been pointed out that Planck and WMAP data are much in better agreement after the application of the Doppler boosting correction 
\cite{Aghanim:2013suk}, that is, the distortion of the CMB anisotropy pattern induced by the proper motion of the observer with respect to CMB rest frame.

In this paper we analyse CMB maps from both WMAP and Planck.
For WMAP we consider three releases of ILC (Internal Linear Combination of the multi-frequency) maps of the CMB sky \cite{Hinshaw:2006ia},
namely we use WMAP ILC 5 year \cite{Hinshaw:2008kr}, WMAP ILC 7 year \cite{Larson:2011} and WMAP ILC 9 \cite{Bennett:2012zja}.
See also \cite{eriksen2004foreground} for further details about the ILC method. 
While for the Planck satellite we use two maps of the 2013 cosmological release of data \cite{Ade:2013hta}:
SMICA, Spectral Matching Independent Component Analysis, \cite{Cardoso:2008qt}, that implements a parametric approach for foreground reduction in the harmonic 
domain\footnote{In fact we consider an inpainted SMICA map which has been produced by replacing the masked pixels with a constrained Gaussian realization obtained by the method described in \cite{inpainting}.}, and
NILC, which employs a spherical needlet version of the ILC algorithm \cite{Delabrouille:2008qd}.

\begin{figure}[h!]
\centering
{\includegraphics[width=12.4cm]{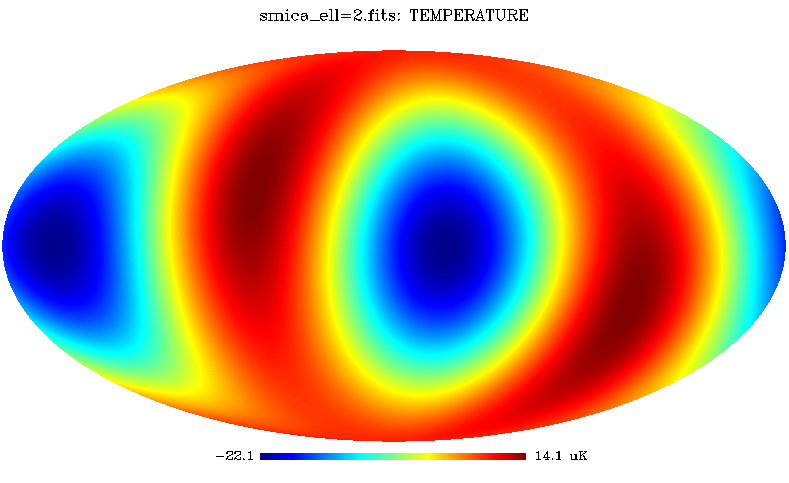}}\hspace{0.56cm} 
{\includegraphics[width=12.4cm]{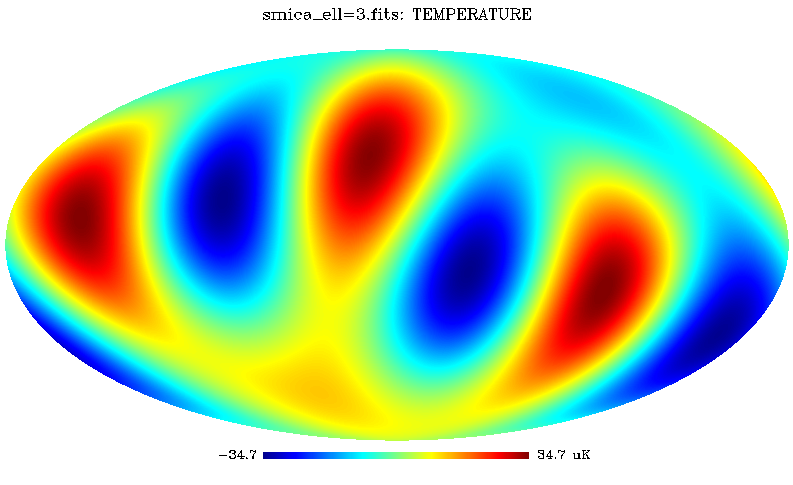}}
\caption{The $\ell=2$ (upper panel) and $\ell=3$ (lower panel) contributions to the Planck SMICA map.}
\label{alignment}
\end{figure}

The alignments are visually illustrated in Fig.~\ref{alignment} where $\ell=2$ and $\ell=3$ of the Planck SMICA map are shown as a representative case.


\subsection{Multipoles vectors}
\label{mv}
It is customary to expand CMB anisotropy maps into spherical harmonics.
However in the context of multipole alignments, it is very convenient to use an alternative and completely equivalent representation, namely multipole (or Maxwell) vectors expansion \cite{Weeks:2004cz,Copi:2003kt,Copi:2013jna}.
The fundamental idea is that the information contained in each set of (complex) $a_{\ell m}$ coefficients for any integer $m=-\ell, ..., \ell$, can be recast 
in $\ell$ unit (real) vectors $\widehat{v}_{i}$ and one (real) amplitude $A^{\ell}$:
\begin{equation}
a_{\ell m} \rightarrow A^{(\ell)}, \widehat{v}_{1}
... \widehat{v}_{\ell} .
\label{almvectors}
\end{equation}
In fact, we note that strictly speaking the term vector is improper here because we should rather speak of axes or directions. 
This happens because the association given in
Eq.~(\ref{almvectors}) is defined up to a ``global'' sign.

The main advantage of this formalism is that it is much easier to build quantities invariant under rotation from multipole vectors rather than from $a_{\ell m}$.
The latter is rather an important point because we will make use in the following of estimators based on rotation invariant quantities.
Unfortunately no closed analytical expression for Eq.~(\ref{almvectors}) is available. Therefore, numerical routines must be used to build the vectors.
Further details and properties can be found in \cite{Weeks:2004cz,Copi:2003kt,Copi:2013jna}.

\subsection{Estimators}
\label{estim}

We build eight estimators, all defined in the interval $[0,1]$ \cite{Gruppuso:2007cn,Copi:2003kt,Copi:2013jna,Abramo:2006gw}.
Of these, six are for the quadrupole/octupole alignment: 
\begin{align} 
&S = \frac{1}{3} \sum_{j=1}^{3} |\widehat{q} \cdot o_{j}| , \label{S} \\
&T = 1 - \frac{1}{3} \sum_{j=1}^{3} (1- |\widehat{q} \cdot o_{j}|)^{2} , \label{T}\\
&S23 = \frac{1}{3} \sum_{j=1}^{3}  |q \cdot o_{j}| , \label{S23}\\
&T23 = 1 - \frac{1}{3} \sum_{j=1}^{3}(1-|q \cdot o_{j}|)^{2} , \label{T23}\\
&\widehat{S}23 = \frac{1}{3} \sum_{j=1}^{3} |\widehat{q} \cdot \widehat{o}_{j}| , \label{S23HAT}\\
&\widehat{T}23 = 1 - \frac{1}{3} \sum_{j=1}^{3} (1-|\widehat{q} \cdot \widehat{o}_{j}|)^{2} \label{T23HAT} ,
\end{align}
and two for the dipole/quadrupole/octupole alignment:
\begin{align} 
&DQO_S = \frac{1}{4} \left( |q \cdot d|+|o_1 \cdot d|+|o_2 \cdot d|+|o_3 \cdot d| \right) , \label{dqo_estiamtor} \\
&DQO_T = 1 - \frac{1}{4} \left[ (1 - |q \cdot d|)^{2} +( 1 - |o_1 \cdot d|)^{2} + ( 1 - |o_2 \cdot d|)^{2} + (1 - |o_3 \cdot d|)^{2} \right]  \label{dqo_estiamtor1}  \, .
\end{align}
In the above equations, the symbol $\hat{\phantom{x}}$ denotes the unit vector, and the area vectors $q$ and $o_j$ are defined via the following vector products:
\begin{align} 
& q = {q}_{21} \times {q}_{22} , \\
& {o}_{1} = {o}_{32} \times {o}_{33} , \\
& {o}_{2} = {o}_{33} \times {o}_{31} , \\
& {o}_{3} = {o}_{31} \times {o}_{32}  ,
\end{align}
where $q_{2j}$ (with $j = 1, 2$) represent the two multipole vectors associated to the quadrupole and 
$o_{3i}$ (with $i = 1, 2, 3$) represent the three multipole vectors associated to the octupole. 
The vector $d$ represents the dipole direction which reads $(l,b)= (263^{\circ} .99,48^{\circ} .26)$ in Galactic coordinates.
Note the presence of the absolute values in the definition of the estimators in Eqs.~(\ref{S})-(\ref{dqo_estiamtor1}) which is due to the fact that multipole vectors define directions, i.e. they are 
headless vectors, see Section \ref{mv}. 

The estimators introduced in Eqs.~(\ref{S})-(\ref{dqo_estiamtor1}) can be divided in ``S'' and ``T'' statistics as denoted by the labels. 
They measure ``distance'' from a situation of complete misalignment, i.e.\ orthogonality, which is associated to zero in both cases,
whereas complete alignment, i.e.\ parallelism, is represented by the value $1$.
However, the ``S'' estimators weight the cosine contributions from the scalar product linearly 
while the ``T'' estimators weight it quadratically. 
Note that in principle these two sets do contain different statistical information but  
we anticipate that they provide very similar results \cite{Copi:2010na}.



\subsection{Simulations pipeline and observed data analysis}
\label{pipeline}

We perform $10^5$ Monte Carlo simulations, extracting $a_{\ell m}$ coefficients from the Planck 2013 $\Lambda$CDM fiducial model\footnote{We have tested that the particular model chosen is irrelevant.}.
For each realization, we transform to multipole vectors 
employing the publicly available code written by Copi et al.
\cite{Copi:2003kt}, whose use is acknowledged here\footnote{See http://www.phys.cwru.edu/projects/mpvectors/ }. 
Then, for each of the performed realizations we compute the eight estimators defined in Eqs.~(\ref{S})-(\ref{dqo_estiamtor1}). 
We therefore can build the empirical distributions of the estimators in the $\Lambda$CDM model, see green histograms in Fig.~\ref{plot}. 
For the dipolar model, our pipeline flows in a similar way. The only difference is that once the $a_{\ell m}$ are drawn, 
we transform them to a real space map, i.e. $\Delta T / T|_{iso}$, and use Eq.~(\ref{dipolarmodel}) to compute $\Delta T / T |_{mod}$. 
We then go back to harmonic space, i.e. $$\Delta T / T |_{mod} \rightarrow a_{\ell m}^{mod} \, , $$ and use these $a_{\ell m}^{mod}$ to compute the multipole vectors. 
Once this is repeated $10^5$ times, we can build the eight empirical distributions of the considered estimators in the dipolar model, see the red histograms in Fig.~\ref{plot}.

Of course the same estimators are evaluated for five observed CMB maps, see Section \ref{data}.
These values are represented by the vertical lines in Fig.~\ref{plot}: WMAP ILC 5 in blue, WMAP ILC 7 in pink, WMAP ILC 9 in balck, Planck 2013 NILC in cyan and Planck 2013 SMICA in magenta.
In fact before evaluating these numbers, we have applied a ``boost correction'' to the observed $a_{\ell m}$ coefficients.
This is necessary because the observed quadrupole is slightly affected by the motion of the satellite with respect to the CMB rest frame.
The details of this correction for every multipole $\ell$ are given in the next subsection.

\subsubsection{Boost correction}
\label{boost}

It is possible to show, see e.g. \cite{Kosowsky:2010jm,Amendola:2010ty}, that the spherical harmonic coefficients, $a_{\ell m}^{RF}$, observed in the CMB rest frame (hereafter $S_{cmb}$) are related to 
the spherical harmonic coefficients, $a_{\ell m}^{\prime}$ defined in a frame $S$ which is moving in the $\hat{z}$ direction at velocity $v$ with respect to $S_{cmb}$, in the following way
\begin{equation}
a_{\ell m}^{\prime} = \sum_{\ell'=0}^{\mathcal{1}} a^{RF}_{\ell' m} I^{m}_{\ell' \ell}(v) \, ,
\label{boosting}
\end{equation}
where no sum on $m$ is understood and where the $I^{m}_{\ell' \ell}(v)$ is defined as
\begin{equation}
I^{m}_{\ell' \ell}(v)=\int_{-1}^{+1} 2 \pi \frac{\sqrt{1-v^{2}}}{1+vx} \widetilde{P}^{m}_{\ell'}(x) \widetilde{P}^{m}_{\ell} \left( \frac{x + v}{1 + vx} \right) dx ,
\end{equation}
with the $\widetilde{P}_{\ell}^{m}$ functions defined through the Legendre polynomial $P_{\ell}^{m}$ as
\begin{equation}
\widetilde{P}^{m}_{\ell}= \sqrt{ \frac{2 \ell +1}{4 \pi} \frac{(\ell-m)!}{(\ell+1)!}} P^{m}_{\ell}  \, .
\end{equation}
In fact we need to invert Eq.~(\ref{boosting}) and ``deboost'' the WMAP and Planck observations.
This can be done using the following orthonormality relation
\begin{equation}
\sum _{\ell'} I^{m}_{\ell' \ell_{1}} I^{m}_{\ell' \ell_{2}} = \delta _{\ell_{1} \ell_{2}} \, ,
\end{equation} 
and considering that 
\begin{equation}
I^{m}_{\ell' \ell}(v) = I^{m}_{\ell \ell'}(-v) \, .
\end{equation} 
Therefore one finds
\begin{equation}
a^{RF}_{\ell m} = \sum_{\ell'} a^{\prime}_{\ell' m} I_{\ell' \ell}^{m}(-v) \, .
\label{alm_deboosting}
\end{equation}
In practice, only $\ell=2$ has to be corrected by this kinematic term. For this multipole, the typical correction is roughly around $10-30 \%$. 
For $\ell \ge 3$ this effect is completely negligible. For the octupole the maximum deviation is computed to be of the order of $0.1\%$.
See Appendix \ref{appendix} where explicit values are reported.

\section{Results}
\label{4}

\begin{figure}[h!]
\centering
{\includegraphics[width=3.4cm]{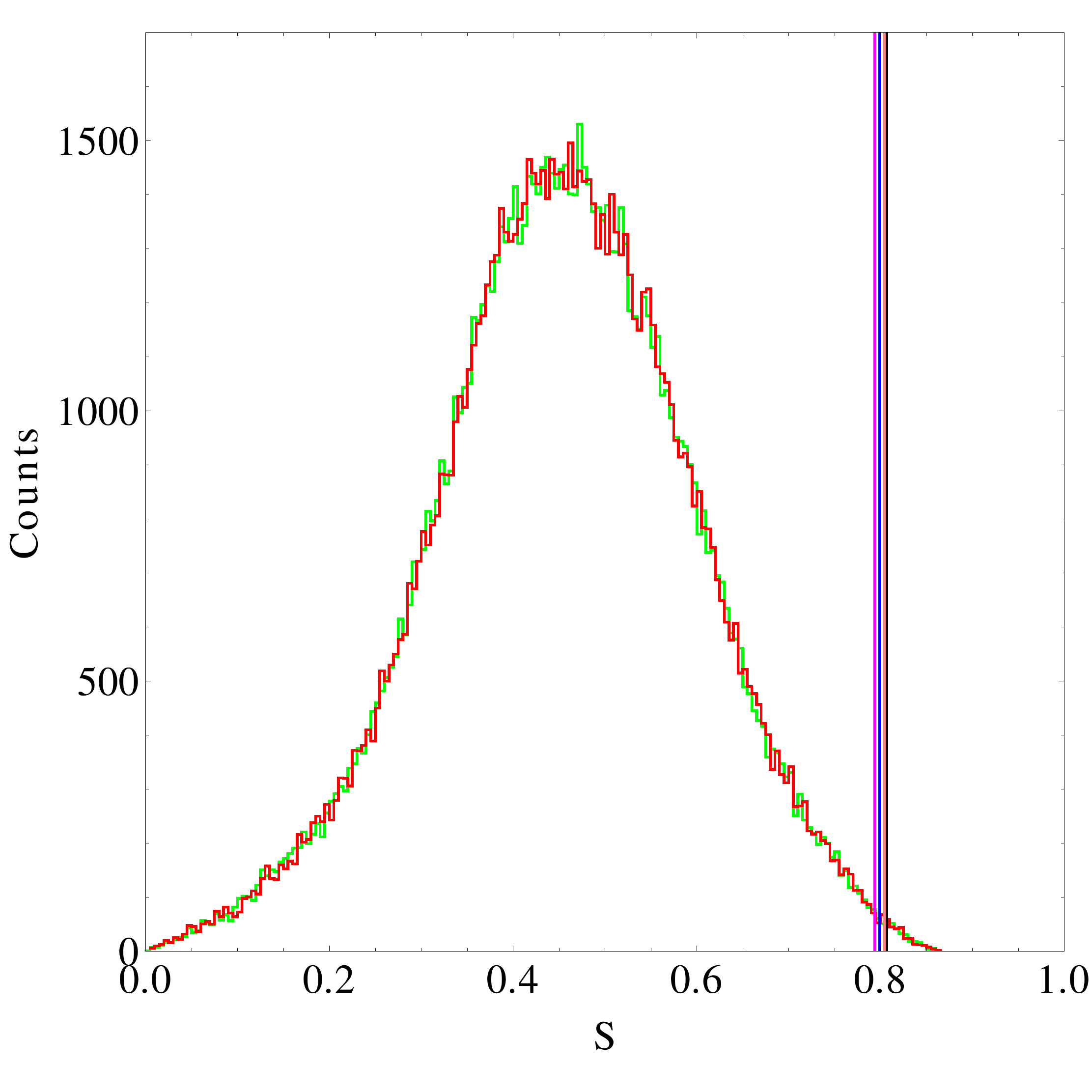}}\hspace{0.01cm}
{\includegraphics[width=3.4cm]{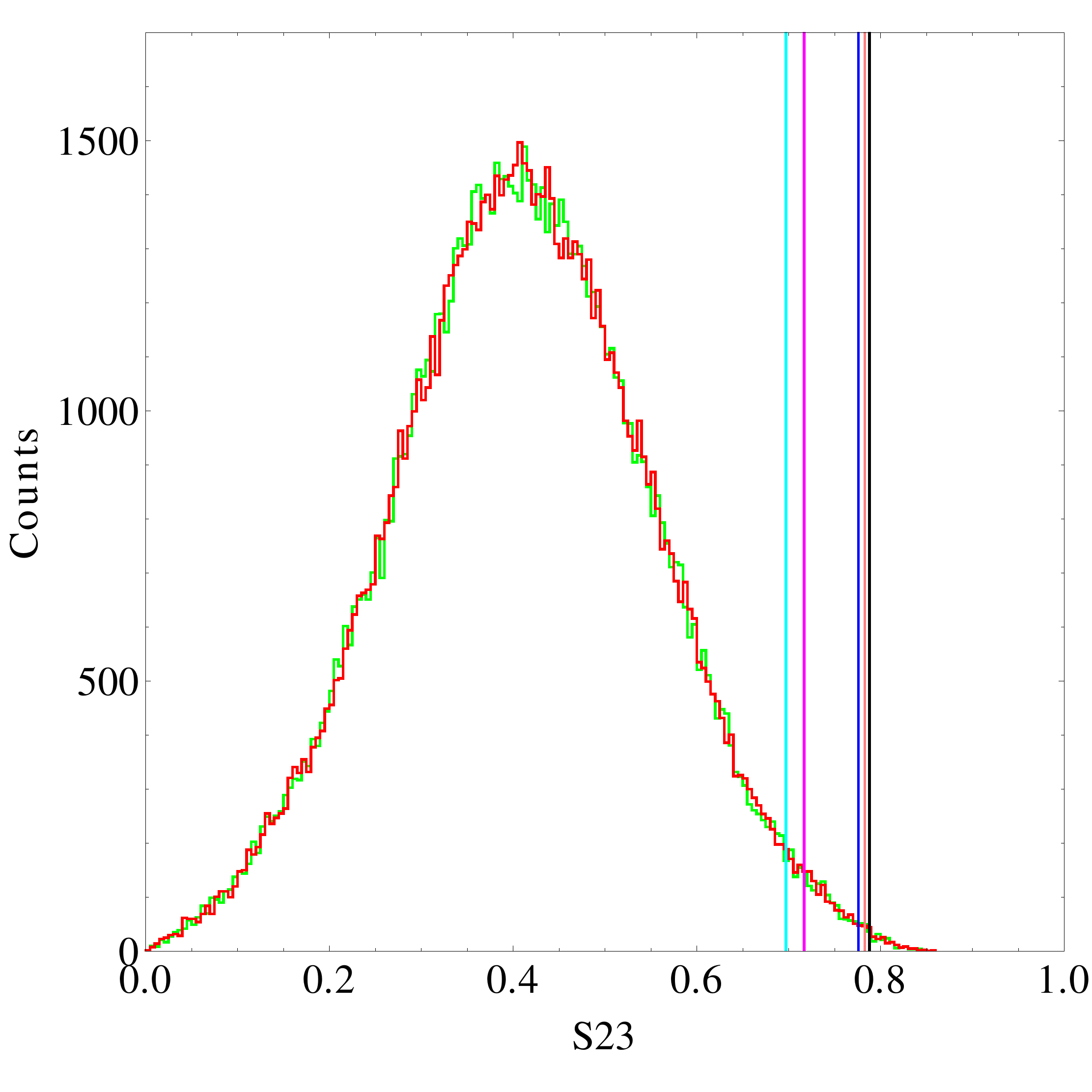}}\hspace{0.01cm}
{\includegraphics[width=3.4cm]{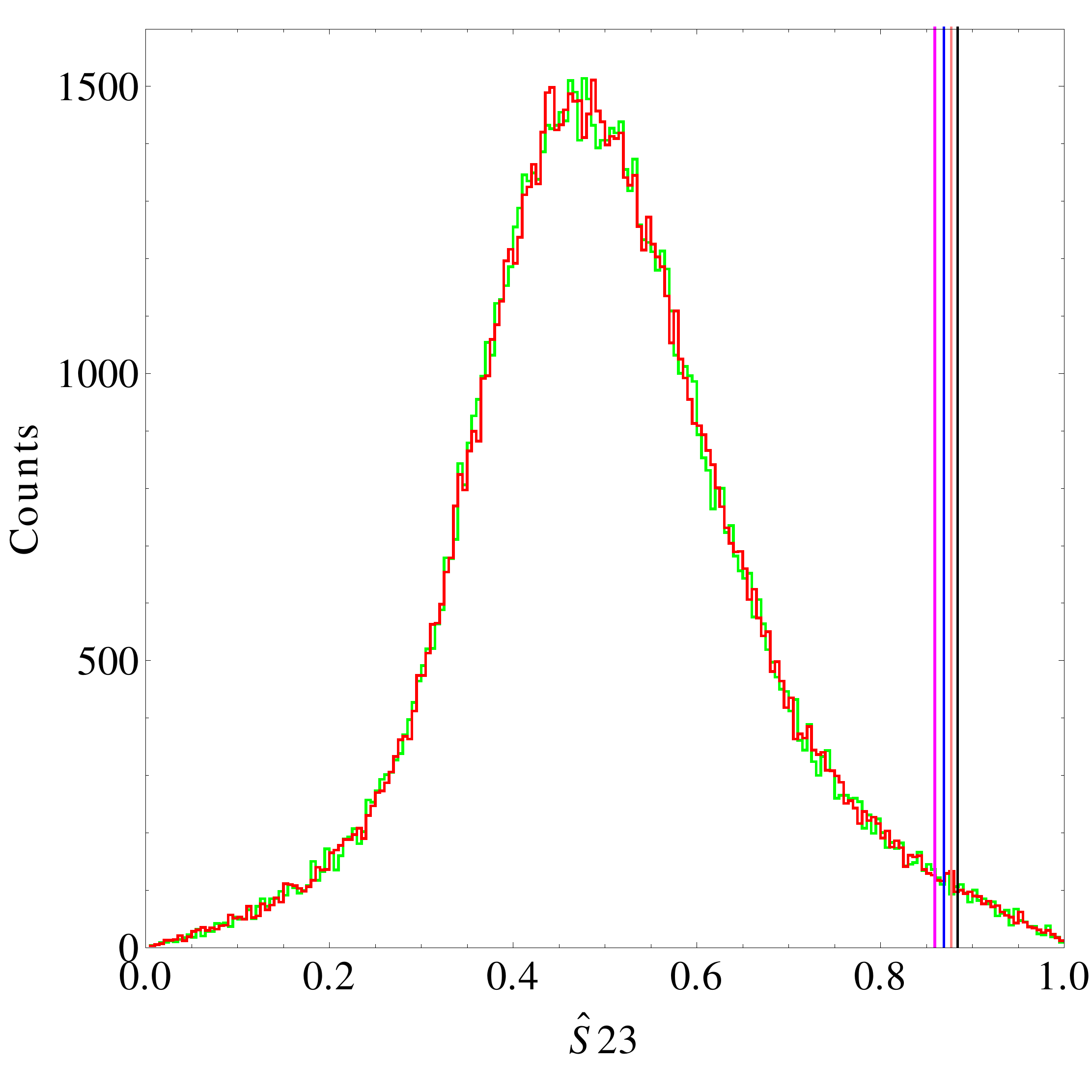}}\hspace{0.01cm}
{\includegraphics[width=3.4cm]{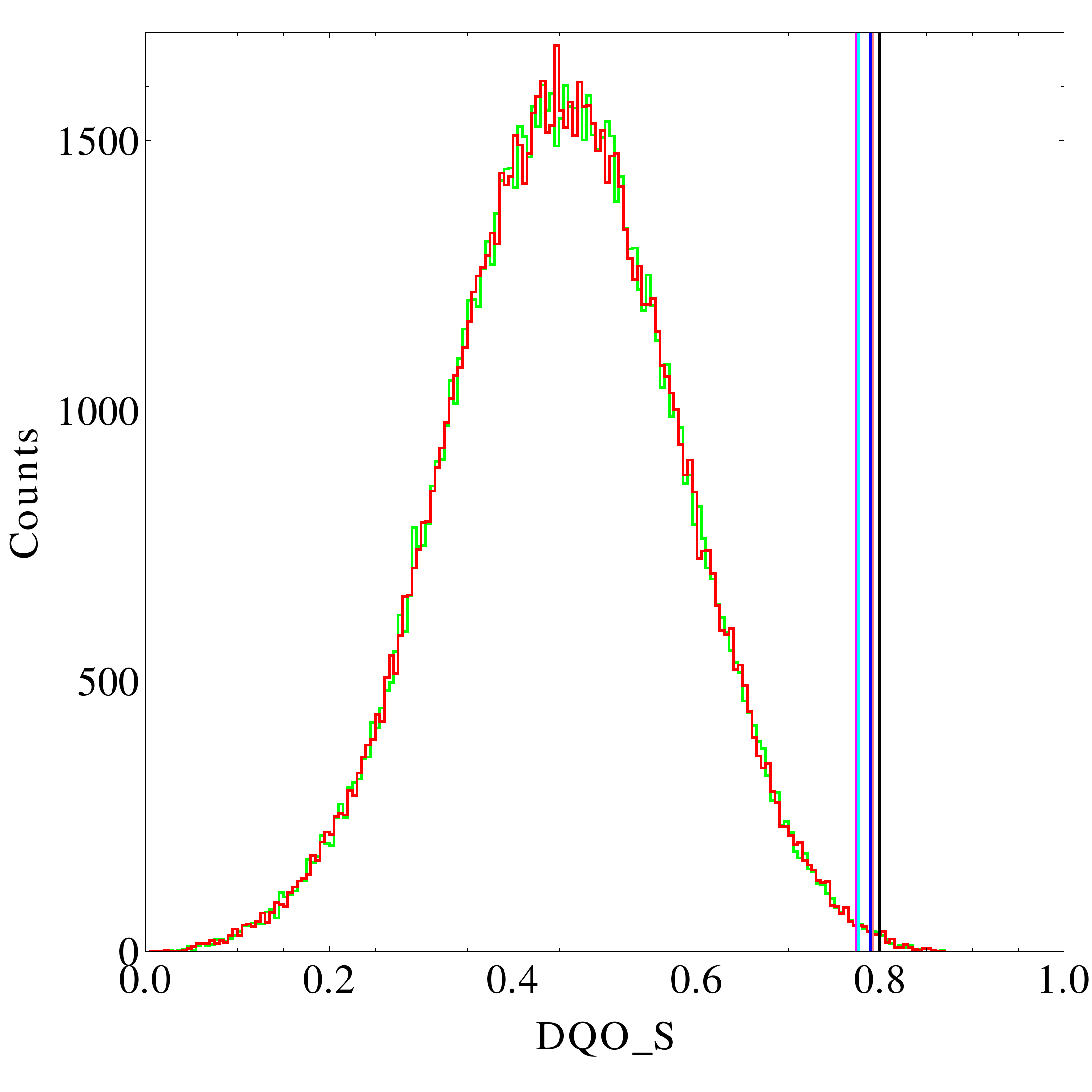}}\hspace{0.01cm}

{\includegraphics[width=3.4cm]{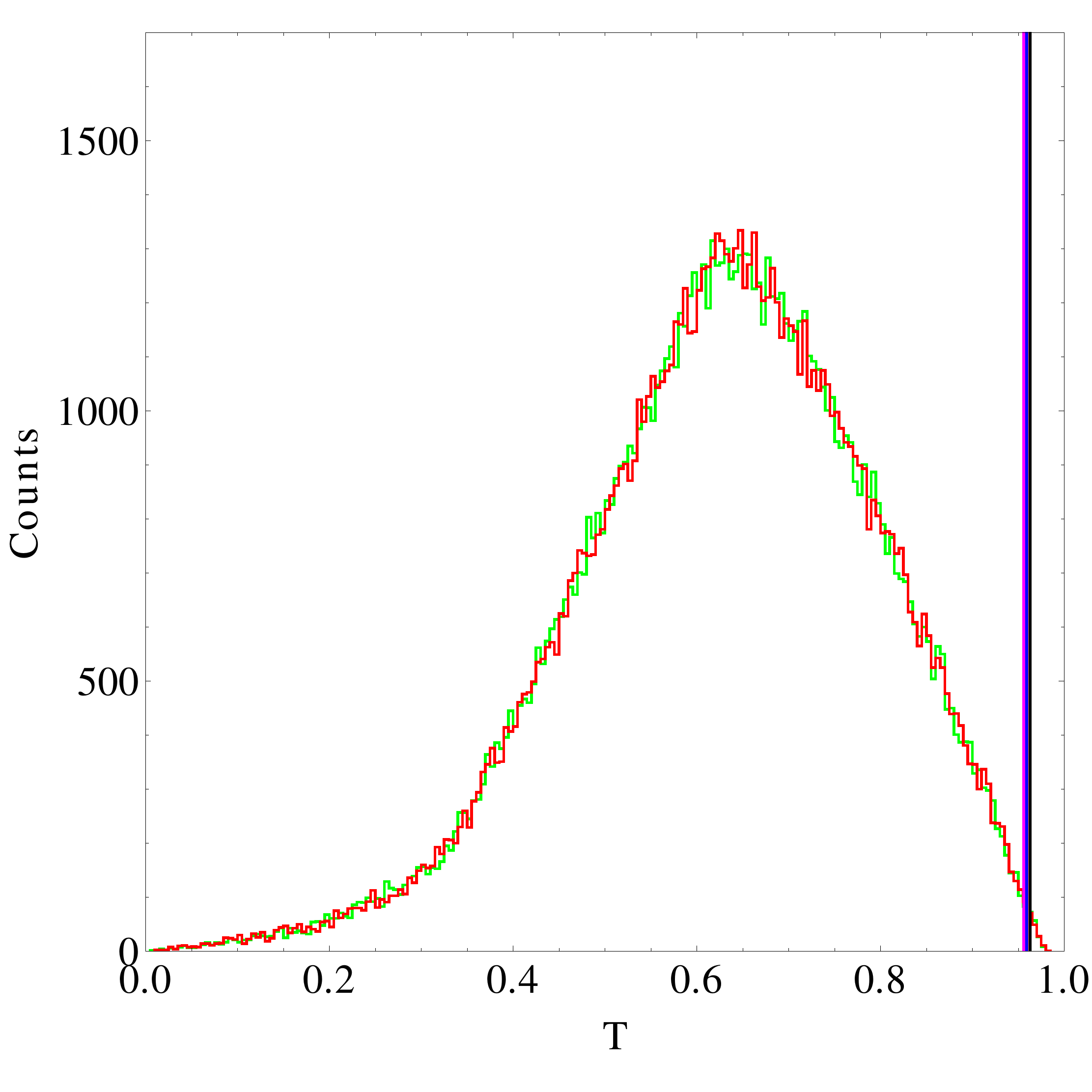}}\hspace{0.01cm}
{\includegraphics[width=3.4cm]{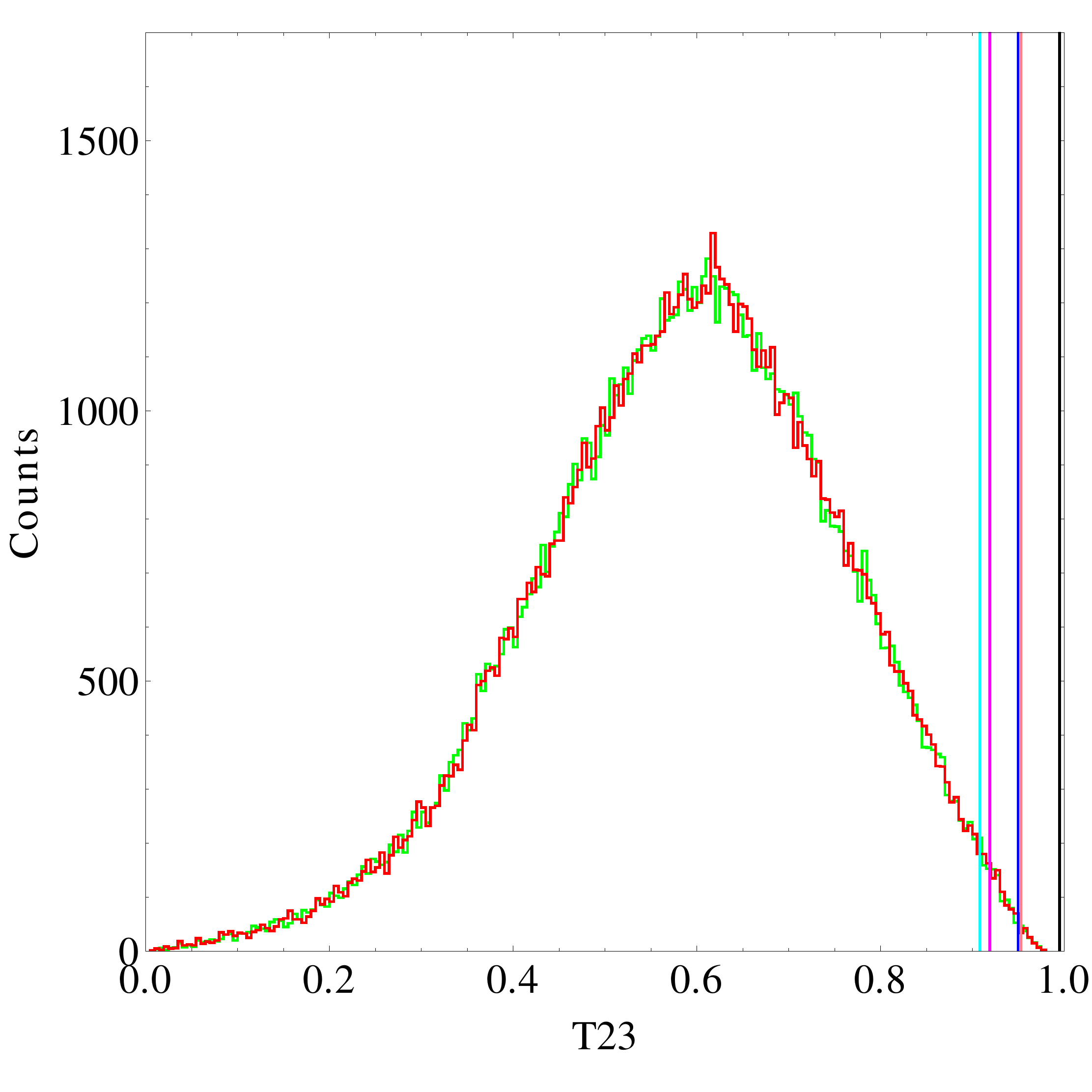}}\hspace{0.01cm}
{\includegraphics[width=3.4cm]{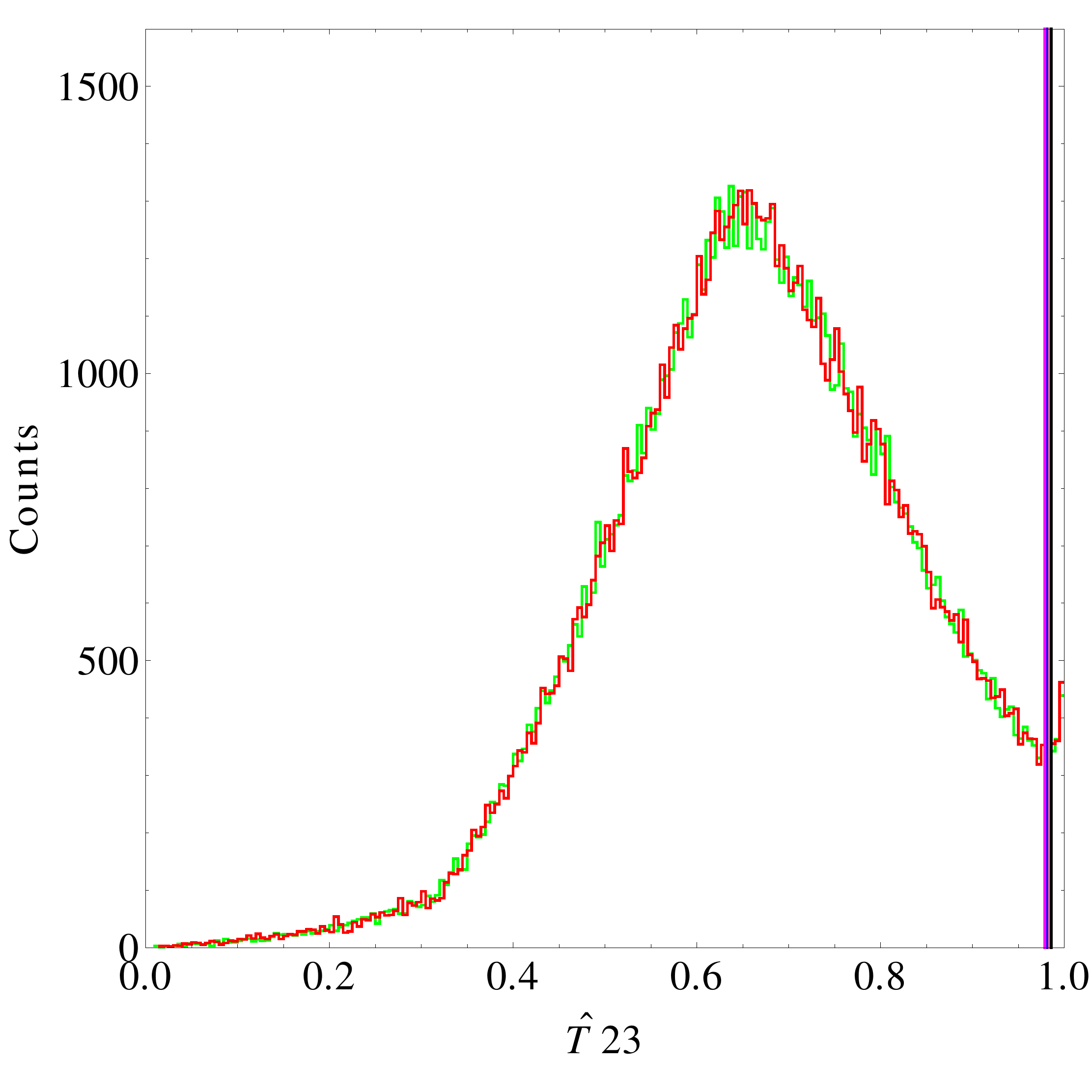}}\hspace{0.01cm}
{\includegraphics[width=3.4cm]{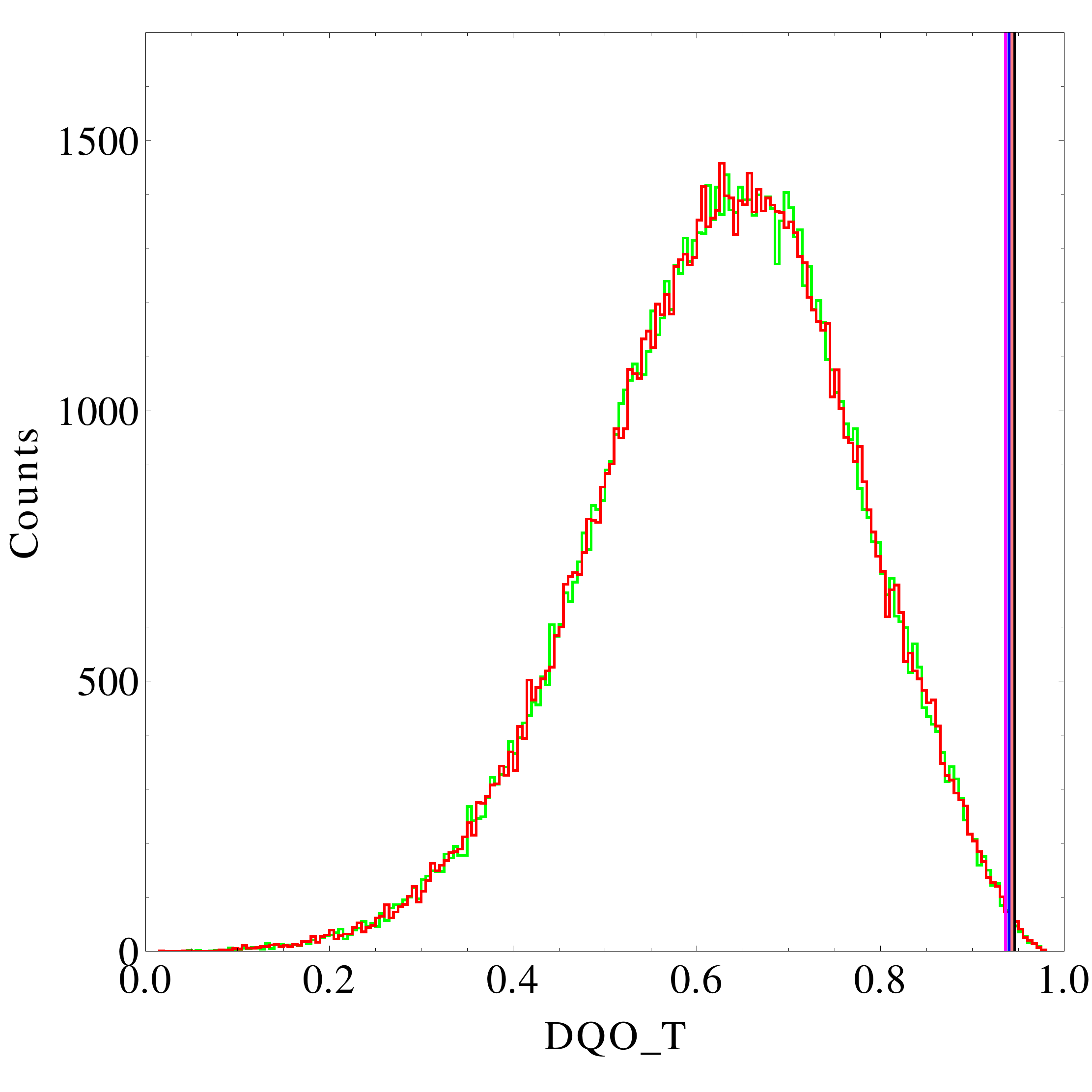}}\hspace{0.01cm}
\caption{ S statistic for the upper row and T statistic for the lower row. 
Green histograms for the empirical distribution of the considered estimators in $\Lambda$CDM and red for the dipolar model.
From left to right we consider $S$, $S23$, $\widehat{S}23$ and $DQO_S$ in the first row and similarly
$T$, $T23$, $\widehat{T}23$ and $DQO_T$ in the second row. Vertical lines are for the observed estimators (already boost-corrected):
WMAP ILC 5 in blue, WMAP ILC 7 in pink, WMAP ILC 9 in balck, Planck 2013 NILC in cyan and Planck 2013 SMICA in magenta. 
In each panel we show the counts in the y-axis and the estimator in the x-axis.
}
\label{plot}
\end{figure}

Our results are shown in Fig.~\ref{plot}.
We evaluate the level of anomaly comparing the histograms with the observed values, i.e. the vertical bars in Fig.~\ref{plot}.
We consider both the $\Lambda$CDM and dipolar, and for each analyzed CMB map,
i.e. WMAP ILC 5, WMAP ILC 7, WMAP ILC 9, Planck 2013 NILC and Planck 2013 SMICA. 

At the price of a slight inaccuracy in terminology, we define the 
probability to exceed, henceforth PTE, as the number of the simulated counts that have the value of the considered estimator
smaller that the observed value. These values are reported in Table \ref{estimator} and the PTEs are provided in Table \ref{allineamento_qo} and 
in Table \ref{allineamento_dqo}.

\begin{table}[h!]
\tiny
\caption{Values of the estimators extracted from the WMAP and Planck CMB maps.}
\medskip
\begin{tabular}{cccccc}
\hline \bigstrut[t] \\
\multicolumn{1}{c}{\textbf{Estimator}} & \multicolumn{1}{c}{\textbf{WMAP ILC 5 yr}} & \multicolumn{1}{c}{\textbf{WMAP ILC 7 yr}} & \multicolumn{1}{c}{\textbf{WMAP ILC 9 yr}} & \multicolumn{1}{c}{\textbf{Planck SMICA}} & \multicolumn{1}{c}{\textbf{Planck NILC}} 
\bigstrut[t]  \\
\bigstrut[t] \\ \hline \\
\bigstrut[t] \\
\multicolumn{1}{c}{\textbf{S}} & 0.799 & 0.804 & 0.807 & 0.794 & \multicolumn{1}{c}{0.804} \\ \bigstrut[t] \\
\multicolumn{1}{c}{\textbf{T}} & 0.959 & 0.962 & 0.963 & 0.956 & \multicolumn{1}{c}{0.962} \\ \bigstrut[t] \\
\multicolumn{1}{c}{\textbf{S23}} & 0.776  & 0.783 & 0.788 & 0.718  & \multicolumn{1}{c}{0.697} \\ \bigstrut[t] \\
\multicolumn{1}{c}{\textbf{T23}} & 0.949  & 0.953 & 0.955 & 0.919  & \multicolumn{1}{c}{0.908} \\ \bigstrut[t] \\
\multicolumn{1}{c}{$\mathbf{\widehat{S}23}$} & 0.869 & 0.877 & 0.884 & 0.859 & \multicolumn{1}{c}{0.877} \\ \bigstrut[t] \\
\multicolumn{1}{c}{$\mathbf{\widehat{T}23}$} & 0.982 & 0.984 & 0.986 & 0.979 & \multicolumn{1}{c}{0.985} \\ \bigstrut[t]  \\
\multicolumn{1}{c}{$ \mathbf{DQO_S}$} & 0.789 & 0.792 & 0.799 & 0.774 & \multicolumn{1}{c}{0.776} \\ \bigstrut[t] \\
\multicolumn{1}{c}{$ \mathbf{DQO_T}$} & 0.940 & 0.943 & 0.946 & 0.936 & \multicolumn{1}{c}{0.944} \\ \bigstrut[t]  \\ \hline
\end{tabular}
\label{estimator}
\end{table}

%

\begin{table}[h!]
\tiny
\caption{Percentage of anomaly for the quadrupole/octupole alignment, for all analysed estimators ($S$, $T$, $S23$, $T23$, $\widehat{S}23$ and $\widehat{T}23$) for the WMAP data (WMAP ILC 5 yr, WMAP ILC 7 yr and WMAP ILC 9 yr) and for the Planck data (Planck SMICA and Planck NILC). }
\medskip
\begin{tabular}{ccccccccccc}
\hline \bigstrut[t] \\
\multicolumn{1}{c}{} & \multicolumn{2}{c}{\textbf{WMAP ILC 5 yr}} & \multicolumn{2}{c}{\textbf{WMAP ILC 7 yr}} & \multicolumn{2}{c}{\textbf{WMAP ILC 9 yr}} & \multicolumn{2}{c}{\textbf{Planck SMICA}} &
\multicolumn{2}{c}{\textbf{Planck NILC}} \\
\bigstrut[t] \\
\multicolumn{1}{c}{\textbf{Estimator}} & \textbf{$\Lambda$CDM} & \textbf{Dipolar} & \textbf{$\Lambda$CDM} & \textbf{Dipolar} & \textbf{$\Lambda$CDM} & \textbf{Dipolar} & \textbf{$\Lambda$CDM} & \textbf{Dipolar} & \textbf{$\Lambda$CDM} & \multicolumn{1}{c}{\textbf{Dipolar}} \bigstrut[t]  \\
\bigstrut[t] \\ \hline \\
\bigstrut[t] \\
\multicolumn{1}{c}{\textbf{S}} & 99.647 & 99.640 & 99.701 & 99.701 & 99.750 & 99.731 & 99.581 & 99.578 & 99.704 & \multicolumn{1}{c}{ 99.707} \\ \bigstrut[t] \\
\multicolumn{1}{c}{\textbf{T}} & 99.828 & 99.832 & 99.856 & 99.866 & 99.873 & 99.880 & 99.775 & 99.769 & 99.856 & \multicolumn{1}{c}{99.866} \\ \bigstrut[t] \\
\multicolumn{1}{c}{\textbf{S23}} & 99.722  & 99.724 & 99.793 & 99.791 &  99.838 & 99.830 & 98.649 & 98.606 & 97.990 & \multicolumn{1}{c}{97.951} \\ \bigstrut[t] \\
\multicolumn{1}{c}{\textbf{T23}} & 99.863  & 99.868 & 99.892 & 99.891 & 99.905 & 99.906 & 99.217 & 99.207 & 98.861 & \multicolumn{1}{c}{98.833} \\ \bigstrut[t] \\
\multicolumn{1}{c}{$\mathbf{\widehat{S}23}$} & 98.355  & 98.308 & 98.569 & 98.539 & 98.689 & 98.676 & 98.128 & 98.089 & 98.550 & \multicolumn{1}{c}{98.523} \\ \bigstrut[t] \\
\multicolumn{1}{c}{$\mathbf{\widehat{T}23}$} & 98.654 & 98.646 & 98.839 & 98.802 & 98.901 & 98.881 & 98.420 & 98.379 & 98.839 & \multicolumn{1}{c}{98.806} \\ \bigstrut[t]  \\ \hline
\end{tabular}
\label{allineamento_qo}
\end{table}

\begin{table}[h!]
\tiny
\caption{Percentage of anomaly for the dipole/quadrupole/octupole alignment, for all analysed estimators (DQO$ \_$S and DQO$ \_$T) for the WMAP data (WMAP ILC 5 yr, WMAP ILC 7 yr and WMAP ILC 9 yr) and for the Planck data (Planck SMICA and Planck NILC). }
\medskip
\begin{tabular}{ccccccccccc}
\\ \hline
\bigstrut[t] \\
\multicolumn{1}{c}{} & \multicolumn{2}{c}{\textbf{WMAP ILC 5 yr}} & \multicolumn{2}{c}{\textbf{WMAP ILC 7 yr}} & \multicolumn{2}{c}{\textbf{WMAP ILC 9 yr}} & \multicolumn{2}{c}{\textbf{Planck SMICA}} &
\multicolumn{2}{c}{\textbf{Planck NILC}} \\
\bigstrut[t] \\
\multicolumn{1}{c}{\textbf{Estimator}} & \textbf{$\Lambda$CDM} & \textbf{Dipolar} & \textbf{$\Lambda$CDM} & \textbf{Dipolar} & \textbf{$\Lambda$CDM} & \textbf{Dipolar} & \textbf{$\Lambda$CDM} & \textbf{Dipolar} & \textbf{$\Lambda$CDM} & \multicolumn{1}{c}{\textbf{Dipolar}} \bigstrut[t]  \\
\bigstrut[t] \\ \hline \\
\bigstrut[t] \\
\multicolumn{1}{c}{\textbf{DQO$ \_ $S}} & 99.803 & 99.796 & 99.829 & 99.823  & 99.872 & 99.865 & 99.672 & 99.662 & 99.687 & \multicolumn{1}{c}{99.681} \\ \bigstrut[t] \\
\multicolumn{1}{c}{\textbf{DQO$ \_ $T}} & 99.776 & 99.779 & 99.810 & 99.808 & 99.859 & 99.851 & 99.725 & 99.728 & 99.825 & 
\multicolumn{1}{c}{99.823} \\ 
\bigstrut[t]  \\ \hline
\end{tabular}
\label{allineamento_dqo}
\end{table}
\newpage

A few comments are in order.
First, the empirical histograms for $\Lambda$CDM and dipolar model are very similar. 
This means it is not easy to distinguish between the two models on basis of the observed alignments.
Second, all vertical lines are very close to each other. This means that at large angular scale in temperature the CMB maps 
obtained with two different experiments and with three different methods are very similar in terms of phases.
Third, all vertical bars, for all the considered estimators, stand in the right hand part of the histograms. 
This means that data tend to show alignments of the considered low multipoles.
The significance of these alignments is in general larger than $99\%$, with few cases at the level of $98-99\%$, and can be as large as $99.9\%$ in selected cases.

\section{Conclusion}
\label{conclusion}

We have tested the CMB quadrupole/octupole and dipole/quadrupole/octupole alignments for several foreground cleaned products for both WMAP (5, 7 and 9 year data) as well as Planck 2013 data.
Specifically, we have considered the WMAP ILC products for the several year releases and Planck NILC and SMICA maps. We have used a total of eight estimators based on the multipole vector formalism, two for the dipole/quadrupole/octupole and six for quadrupole/octupole alignments. All these estimators are supported by a large Monte Carlo of $10^5$ independent maps. We report that all the data combinations and all the estimators we have tested  exhibit anomalous alignments for both combinations of multipoles considered, typically at the $98\%$-$99\%$ level, and up to $99.9\%$ in selected cases. The consistent pattern for the alignments observed in both WMAP and Planck strongly disfavours an origin of the effect related to unaccounted instrumental systematics. The wide frequency leverage of the Planck data (30 to 353 GHz), weakens considerably the case for residual foreground emission. The fact that we find consistent results also among different foreground separation procedures (SMICA, NILC and WMAP's ILC) makes this conclusion stronger. We have also investigated the possibility that the phenomenological dipolar model may provide a better framework for the existence of the observed alignments with respect to plain $\Lambda$CDM. This possibility is, in principle, intriguing because the dipolar model has gathered some success in accounting for other anomalies, e.g. the hemispherical asymmetry. We report negative findings on this last issue: the dipolar model does not seem to be able to accomodate for the existence of anomalies significantly better than $\Lambda$CDM.

\section*{Acknowledgments}

We are grateful to Bruce Partridge for valuable comments.
We acknowledge the use of the publicly available code for the multipole vectors decomposition (\url{http://www.phys.cwru.edu/projects/mpvectors/}) described in \cite{Copi:2003kt}. 
We also acknowledge the use of the HEALPix package (\url{http://healpix.sourcef orge.net}), see \cite{Gorski:2004by}. 
Some results presented in this papers are based on observations obtained with Planck (\url{http://www.esa.int/Planck}), an ESA science mission with instruments and contributions directly funded by ESA Member States, NASA, and Canada. Moreover, we acknowledge the use of the Legacy Archive for Microwave Background Data Analysis (LAMBDA), part of the High Energy Astrophysics Science Archive Center (HEASARC). 
HEASARC/LAMBDA is a service of the Astrophysics Science Division at the NASA Goddard Space Flight Center.
Work supported by ASI through ASI/INAF Agreement I/072/09/0 for the Planck LFI Activity of Phase E2.

\appendix

\section{Impact of the boost correction}
\label{appendix}

In Table \ref{tab_senza_deb_ell=2_e_ell=3} and \ref{tab_deb__ell=2_e_ell=3} we report the $a_{\ell m}$ for quadrupole and octupole without and with de-boosting correction respectively.
\begin{table}[h!]
\tiny
\caption{$a_{\ell m}$ for $\ell=2$ and $\ell=3$ (no correction applied). Units: $\mu$K.}
\medskip
\begin{tabular}{cccccccc}
\hline \bigstrut[t] \\
\multicolumn{1}{c}{\textbf{ }} & \textbf{$\mathbf{a_{20}}$} & \textbf{$\mathbf{a_{21}}$} & \textbf{$\mathbf{a_{22}}$} & \textbf{$\mathbf{a_{30}}$} & \textbf{$\mathbf{a_{31}}$} & \textbf{$\mathbf{a_{32}}$} & \textbf{$\mathbf{a_{33}}$} \bigstrut[t]  \\
\bigstrut[t] \\ \hline \\
\bigstrut[t] \\
\multicolumn{1}{c}{\textbf{WMAP ILC 5 yr}} & 12.350 & -1.087+6.069i & -14.211-17.858i & 
-6.449 & -12.733+2.443i & 22.019+0.698i & -11.813+33.393i \\ \bigstrut[t] \\
\multicolumn{1}{c}{\textbf{WMAP ILC 7 yr}} & 11.771 & -0.771+6.215i & -14.120-17.941i & 
-6.479 & -12.191+2.026i & 21.999+0.591i & -11.709+33.554i \\ \bigstrut[t] \\
\multicolumn{1}{c}{\textbf{WMAP ILC 9 yr}} & 12.563 & -1.727+6.209i & -13.846-18.017i &
-6.844 & -11.271+1.581i & 21.857+0.535i & -12.060+32.853i \\ \bigstrut[t] \\
\multicolumn{1}{c}{\textbf{Planck 2013 SMICA}} & 13.089 & -1.530+2.497i & -15.503-17.091i & -5.959 & -12.841+1.671i & 22.086+1.670i & -12.465+29.402i \\ \bigstrut[t] \\
\multicolumn{1}{c}{\textbf{Planck 2013 NILC}} & 13.512 & -1.375+1.722i & -13.564-16.325i & -6.117 & -9.547+1.896i & 22.242+1.875i & -12.914+28.340i  \\ \bigstrut[t] \\
\end{tabular}
\label{tab_senza_deb_ell=2_e_ell=3}
\end{table}

\begin{table}[h!]
\tiny
\caption{De-boosted $a_{\ell m}$ for $\ell=2$ and $\ell=3$. Units: $\mu$K. }
\medskip
\begin{tabular}{cccccccc}
\hline \bigstrut[t] \\
\multicolumn{1}{c}{\textbf{ }} & \textbf{$\mathbf{a_{20}}$} & \textbf{$\mathbf{a_{21}}$} & \textbf{$\mathbf{a_{22}}$} & \textbf{$\mathbf{a_{30}}$} & \textbf{$\mathbf{a_{31}}$} & \textbf{$\mathbf{a_{32}}$} & \textbf{$\mathbf{a_{33}}$} \bigstrut[t]  \\
\bigstrut[t] \\ \hline \\
\bigstrut[t] \\
\multicolumn{1}{c}{\textbf{WMAP ILC 5 yr}} & 10.882 & -1.385+8.716i & -13.076-17.619i & 
-6.458 & -12.750+2.477i & 22.066+0.726i & -11.767+33.351i \\ \bigstrut[t] \\
\multicolumn{1}{c}{\textbf{WMAP ILC 7 yr}} & 10.304 & -1.068+8.861i & -12.985-17.701i & 
-6.486 & -12.209+2.059i & 22.046+0.619i & -11.664+33.513i \\ \bigstrut[t] \\
\multicolumn{1}{c}{\textbf{WMAP ILC 9 yr}} & 11.095 & -2.023+8.854i & -12.710-17.777i & -6.855 & -11.288+1.614i & 21.903+0.562i &  -12.017+32.811i \\ \bigstrut[t] \\
\multicolumn{1}{c}{\textbf{Planck 2013 SMICA}} & 11.622 & -1.830+ 5.143i & -14.363-16.852i & -5.964 &  -12.857+1.709i &  22.139+1.696i & -12.421+29.362i \\ \bigstrut[t] \\
\multicolumn{1}{c}{\textbf{Planck 2013 NILC}} & 12.046 & -1.670+4.368i & -12.423-16.086i & -6.122 & -9.563+1.935i & 22.291+1.900i & -12.873+28.301i \\ \bigstrut[t] \\
\end{tabular}
\label{tab_deb__ell=2_e_ell=3}
\end{table}


\end{document}